%% file: main.tex
\documentclass[conference]{IEEEtran}
\IEEEoverridecommandlockouts
\usepackage{amsmath,amssymb,amsfonts}
\usepackage{algorithmic}
\usepackage{graphicx}
\usepackage{textcomp}
\usepackage{xcolor}
\usepackage{bm}
\usepackage{booktabs}
\usepackage{multirow}
\usepackage{tabularx}
\usepackage{comment}
\usepackage{url}
\usepackage{pifont}
\usepackage{color}
\usepackage{siunitx}

\input{settings.tex}

\newcommand{\modified}[1]{\textcolor{black}{#1}}
\begin{document}

\title{Discriminative Neighborhood Smoothing for Generative Anomalous Sound Detection\\
}

\author{\IEEEauthorblockN{Takuya Fujimura}
\IEEEauthorblockA{\textit{Nagoya University} \\
Nagoya, Japan \\
fujimura.takuya@g.sp.m.is.nagoya-u.ac.jp
}
\and
\IEEEauthorblockN{Keisuke Imoto}
\IEEEauthorblockA{\textit{Doshisha University}\\
Kyoto, Japan \\
keisuke.imoto@ieee.org
}
\and
\IEEEauthorblockN{Tomoki Toda}
\IEEEauthorblockA{\textit{Nagoya University}\\
Nagoya, Japan \\
tomoki@icts.nagoya-u.ac.jp
}
}

\maketitle

\begin{abstract}
We propose discriminative neighborhood smoothing of generative anomaly scores for anomalous sound detection.
While the discriminative approach is known to achieve better performance than generative approaches often, we have found that it sometimes causes significant performance degradation due to the discrepancy between the training and test data, making it less robust than the generative approach.
Our proposed method aims to compensate for the disadvantages of generative and discriminative approaches by combining them.
Generative anomaly scores are smoothed using multiple samples with similar discriminative features to improve the performance of the generative approach in an ensemble manner while keeping its robustness.
Experimental results show that our proposed method greatly improves the original generative method, including absolute improvement of \SI{22}{\percent} in AUC and robustly works, while a discriminative method suffers from the discrepancy.
\end{abstract}



\section{Introduction}
Anomalous sound detection~(ASD) aims to automatically detect mechanical failures from sounds emitted by a target machine~\cite{Koizumi_DCASE2020_01,kawaguchi2021description,dohi2022description}.
This technique is used in machine condition monitoring systems within factories.
As anomalous sound data are rarely observed and unknown failures must also be detected in the ASD task, an ASD system needs to be developed without using labeled data.
This is typically done by constructing an anomaly score calculator using only the normal sounds, where high scores should be assigned to anomalous sounds.


ASD methods are classified into two approaches: generative and discriminative approaches.
The generative approach directly models the audio feature of the normal sounds and calculates the degree of deviation from the model as an anomaly score~\cite{Koizumi_DCASE2020_01,suefusa2020anomalous,dohi2021flow,dohi2022disentangling,guan2023time}.
Autoencoder~(AE)~\cite{Koizumi_DCASE2020_01,suefusa2020anomalous}, normalizing flow~(NF)~\cite{dohi2021flow,dohi2022disentangling}, and Gaussian mixture models~(GMM)~\cite{guan2023time} are popular modeling methods.
AE is trained to reconstruct normal sounds and calculates the anomaly score of an observed sound by its reconstruction error.
NF and GMM are trained to maximize its likelihood for normal sounds and calculate the negative likelihood as the anomaly score.
This generative approach is a proper strategy for ASD when the labeled anomalous data is not available during development; however, it is difficult to construct highly accurate generative models.
For this reason, the generative approach tends to have insufficient detection performance.

\begin{figure}[t]
  \centering
  \centerline{\includegraphics[width=0.99\columnwidth]{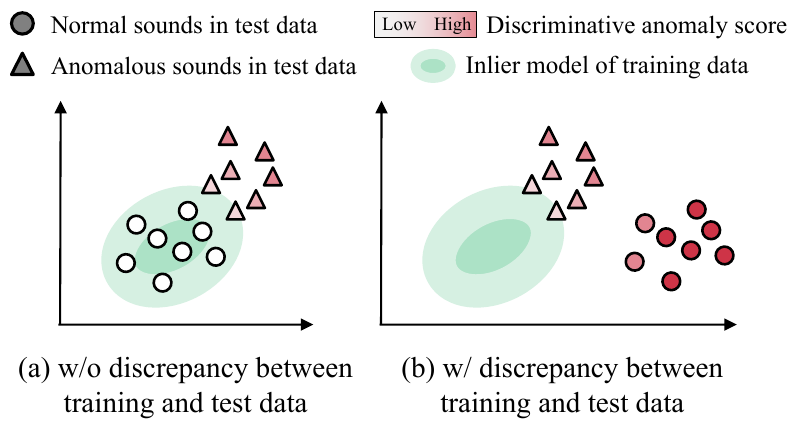}}
  \caption{Illustration of the feature space without and with the discrepancy between the characteristics of the training and test data.}
  \label{fig:motivation}
  \vspace{-8pt}
\end{figure}

Another strategy is the discriminative approach~\cite{Inoue2020,Chen2023,kuroyanagi2022improvement,Primus2020,Wilkinghoff2023}.
This approach acquires the normal area in the discriminative feature space by utilizing the discriminative model trained to classify normal and \textit{pseudo-anomalous} sounds.
The pseudo-anomalous sounds are prepared by various methods such as pitch shifting, time stretching~\cite{Inoue2020}, mixup, changing the statistics of spectrograms~\cite{Chen2023}, and utilizing sounds that are not emitted from a target machine~\cite{kuroyanagi2022improvement,Primus2020}.
This approach assumes that anomalous sounds will be located outside of the normal area (i.e., decrease in the posterior probability of normal sound), and calculates anomaly scores using classification probabilities or deviations from the inlier model constructed with the normal training data~(Fig.~\ref{fig:motivation} (a)).
The discriminative approach has outperformed the generative approach by turning the problem setting from a harder one
(i.e., modeling a probability density function) to easier one (i.e., modeling a posterior probability function).
However, when there is a discrepancy in data characteristics between the training and test data, the test data sometimes deviate from the inlier model even if they are normal sounds~(Fig.~\ref{fig:motivation} (b)).
In this case, the inlier model does not reflect the correct relationship for the anomaly score, leading to significant performance degradation.

The goal of this work is to outperform the generative approach without the risk of performance degradation caused by the discrepancy between the training and test data.
To achieve this goal, we propose the discriminative neighborhood smoothing of a generative method, based on our finding that normal and anomalous sounds tend to be distinguished in the discriminative feature space even in the presence of discrepancies~(Fig.~\ref{fig:motivation} (b)).
The proposed method improves the performance of the original generative approach by ensembling the anomaly scores of multiple samples with similar discriminative features.
In the experimental evaluation, we demonstrate that (1) the proposed method greatly improves the performance of the original generative method and,
(2) robustly works and has the potential to improve its performance even in a situation where the discriminative method does not work due to the discrepancy.
Furthermore, (3) we conduct several analyses to clarify its behavior.

\section{Related Works}

\subsection{Generative approach}
\label{sec:gen}
The autoencoder~(AE)-based method~\cite{Koizumi_DCASE2020_01} is a popular generative method.
This method trains the AE to minimize the following reconstruction loss $\mathcal{L}_{\rm AE}$.
\begin{equation}
\mathcal{L}_{\rm AE}=\frac{1}{MD}\sum_m^M\left\|\psi(\bm{x}^-_m)-f_{\rm G}\left(\psi(\bm{x}^-_m); \bm{\theta}\right)\right\|_2^2,
\end{equation}
where $M$ is a mini-batch size, $\bm{x}^-_m$ is a $m$-th normal sound, $\psi(\cdot)$ extracts $D$-dimensional audio feature such as Mel-spectrogram, $f_{\rm G}$ is a network of AE and $\bm{\theta}$ is the set of its parameters.
During inference, this method calculates the following anomaly score $a_{\rm AE}(\bm{y}_m)$ for the observed sound $\bm{y}_m$.
\begin{equation}
a_{\rm AE}(\bm{y}_m)=\frac{1}{D}\left\|\psi(\bm{y}_m)-f_{\rm G}\left(\psi(\bm{y}_m); \bm{\theta}\right)\right\|_2^2.
\end{equation}
The AE-based method calculates the reconstruction loss for the anomaly score assuming that the reconstruction quality of the anomaly sound will be poor because the network is not trained with anomaly sounds.

Thus, the generative approach, such as the AE-based method, requires highly accurate modeling of normal sounds.
However, this modeling is very challenging, often resulting in insufficient performance in detecting anomalous sounds.

\subsection{Discriminative approach}
\label{sec:dis}
Various discriminative approaches have been investigated~\cite{Inoue2020,Chen2023,kuroyanagi2022improvement,Primus2020,Wilkinghoff2023}, among which SerialOE~\cite{kuroyanagi2022improvement} shows high performance.
It uses sounds emitted from machines other than the target machine as pseudo-anomalous sounds and calculates the anomaly score in the discriminative feature space.
SerialOE is trained in two stages.
First, the discriminative feature extractor $f_{\rm D}(\cdot)$ is trained with the following multi-task loss $\mathcal{L}_{\rm SerialOE}$.
\begin{equation}
    \label{eq:serialoe}
    \begin{split}
        \mathcal{L}_{\rm SerialOE} = \mathcal{L}_{\rm machine}+\lambda_{\rm id} \mathcal{L}_{\rm id},
    \end{split}
\end{equation}
where $\lambda_{\rm id}$ is a hyperparameter for weighting.
$\mathcal{L}_{\rm machine}$ is a loss for classifying normal and pseudo-anomalous sounds as follows:
\begin{equation}
    \mathcal{L}_{\rm machine} = \frac{1}{M}\sum_m^M \mathrm{BCE}\left(t^{\rm mac}_m, \sigma\left(g_{\rm mac}\left(f_{\rm D}\left(\psi\left(\bm{x}_m\right)\right)\right)\right)\right),
\end{equation}
\begin{equation}
    \mathrm{BCE}(p,q)= p \cdot \log q+\left(1-p\right) \cdot \log \left(1-q\right),
\end{equation}
where $\bm{x}_m$ is the $m$-th training data, and $t^{\rm mac}_m$ is the label that is $1$ for the normal sound and $0$ for the pseudo-anomalous sound.
$g_{\rm mac}(\cdot)$ is a function transforming the discriminative feature into a scalar, and $\sigma(\cdot)$ is a sigmoid function.
$\mathcal{L}_{\rm id}$ is a loss for classifying sounds of the target machine in detail using the product ID.
The product ID is an identifier for each individual of the target machine type.
\begin{equation}
    \mathcal{L}_{\rm id} = \frac{1}{MC}\sum_m^M\sum_c^C \mathrm{BCE}\left(t^{\rm id}_{m,c}, \sigma\left(g_{\rm id}\left(f_{\rm D}\left(\psi\left(\bm{x}_m\right)\right)\right)\right)\right),
\end{equation}
where the target machine has $C$ classes for the product ID, $t^{\rm id}_{m,c}$ is its one-hot label.
When $\bm{x}_m$ belongs to the $c$-th class, $t^{\rm id}_{m,c}$ is $1$ and the other elements of $t^{\rm id}_{m}$ are $0$.
This multi-task training manages both major and minor differences in the audio features and constructs a better feature space in which anomalous sounds are distinguished from normal sounds.

Next, an inlier model $h$ such as a GMM is trained with the discriminative features of the normal training data. 
During inference, SerialOE calculates the anomaly score of the observed sound $\bm{y}_m$ using its discriminative feature $f_{\rm D}(\bm{y}_m)$ and a trained inlier model $h$.
For example, GMM can calculate the negative log-likelihood as the anomaly score~(Fig.~\ref{fig:motivation} (a)).

The discriminative approach such as SerialOE has achieved high performance by using the discriminative feature space.
However, when there is a discrepancy between the characteristics of the training and test data, 
the test data forms a different cluster from the cluster of the training data, even if they are normal sounds.
In addition, the anomalous cluster of the test data is sometimes located closer to the inlier model $h$ than the normal cluster of the test data~(Fig.~\ref{fig:motivation} (b)).
In this case, if we use the degree of deviation from $h$ as the anomaly score, the score relationship of normal and anomalous sounds will be reversed, leading to significant performance degradation~\cite{FujimuraNU2023}.
Thus, the discriminative approach has a risk of yielding poor performance due to the discrepancy between the training and test data.


\section{Proposed Method}
\begin{figure}[t]
  \centering
  \centerline{\includegraphics[width=0.99\columnwidth]{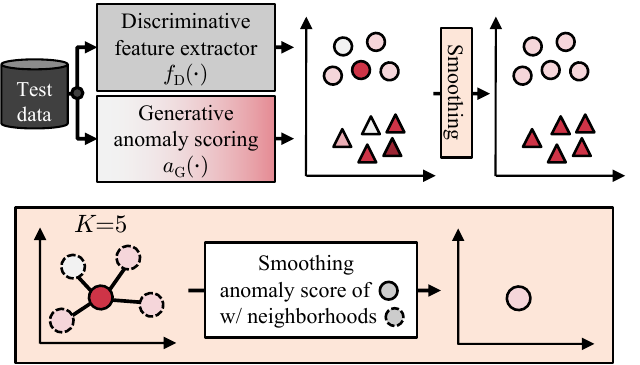}}
  \caption{Overview of the proposed method. Test data are processed with the discriminative feature extractor and generative anomaly scoring in parallel and we obtain the discriminative features to which the generative anomaly scores are assigned.
Anomaly scores are smoothed in the discriminative space and are expected to be modified to the correct score using multiple neighborhood samples.
}
  \label{fig:method}
\end{figure}

We have two important observations:
(1) the generative approach is less susceptible to a discrepancy between the training and test data but its detection performance is insufficient, and
(2) normal and anomalous sounds in the test data can be distinguished in the discriminative feature space even in the presence of the discrepancy~\cite{FujimuraNU2023}.
Based on these observations, we propose the discriminative neighborhood smoothing of the generative approach~(Fig.~\ref{fig:method}).

The proposed method works as follows.
First, we obtain a set of discriminative features $\mathcal{F}=\{f_{\rm D}(\bm{y}_i)\, |\, \bm{y}_i \in \mathcal{Y} \}$ from a set of observed sounds during the operating phase $\mathcal{Y}=\{\bm{y}_0,...,\bm{y}_N\}$ (i.e., $\mathcal{Y}$ is a test dataset).
Next, the proposed method calculates the anomaly score $a_{\rm prop}(\bm{y}_i)$ for the observation $\bm{y}_i$ as follows:
\begin{align}
a_{\rm prop}(\bm{y}_i)=\frac{1}{K}\left(a_{\rm G}(\bm{y}_i)+\sum_{k\in\mathcal{N}_{\mathcal{F}}(\bm{y}_i)}a_{\rm G}(\bm{y}_k)\right),
\end{align}
where $a_{\rm G}(\cdot)$ calculates the anomaly score by some generative method.
$\mathcal{N}_{\mathcal{F}}(\bm{y}_i)$ is the set of indices of neighborhood of $f_{\rm D}(\bm{y}_i)$ on the $\mathcal{F}$ except for $i$, $K$ is the number of neighbors, and $|\mathcal{N}_{\mathcal{F}}(\bm{y}_i)|=K-1$.

The proposed method improves the detection performance from the original generative method by ensembling anomaly scores of multiple samples with similar discriminative features.
For example, a high anomaly score wrongly assigned to normal sound will be modified to a correctly low score by smoothing if most of its neighborhoods are expected to be normal sounds~(Fig.~\ref{fig:method}).
Also, the proposed method is robust to discrepancies because it does not use the deviation from the inlier model $h$ as the anomaly score.
Thus, the proposed method achieves robust performance by relaxing the requirements for the discriminative feature space from both (i) normal and anomalous sounds are distinguished and (ii) normal sounds of test and training data are matched, to only (i).
\modified{It should be noted that the proposed method uses the test data to modify anomaly scores.}
Therefore, the ASD system using our proposed method works by regularly updating $\mathcal{F}$ with the data observed during the operating phase.

\section{Experimental Evaluation}
\subsection{Experimental Setups}
We conducted an experimental evaluation using the DCASE 2021 Task~2 Challenge datasets (MIMII DUE~\cite{Tanabe_WASPAA2021_01} and ToyADMOS2~\cite{Harada2021}).
The datasets included seven machine types: fan, gearbox, pump, slider, ToyCar, ToyTrain, and valve.
It also provided two domains~(source and target domains) and six sections for each machine type to handle the domain-shifted problem, where the section was a subset for the evaluation and it reflected the product ID.
The training data had approximately 1,003 samples of normal data for each section of each machine type, of which only three samples were in the target domain and the others were in the source domain.
The test data had 100 samples of normal and anomalous data for each machine, section, and domain.
Each recording was a 10-second single channel segment sampled at \SI{16}{\kHz} and we standardized its amplitude to have a mean of 0.0 and a variance of 1.0.
Sections 0, 1, and 2 of the test data were used as the validation data to determine the hyperparameters, while Sections 3, 4, and 5 were used as the evaluation data.
To train the feature extractor and the generative method, we used \SI{85}{\percent} of the source domain data and two samples of the target domain data.
The remaining samples were treated as the validation set during training.


Here, we used an AE as $a_G(\cdot)$.
The network was a ten-layer fully connected network~\cite{Koizumi_DCASE2020_01}.
The audio input feature $\psi(\cdot)$ was the five adjacent frames of the log Mel power spectrogram with a window size of \SI{64}{\ms}, a hop size of \SI{32}{\ms}, and 128 Mel-spaced frequency bins in the range of \SIrange{50}{7800}{\Hz}.
We trained the network for 1,500 epochs using the Adam optimizer~\cite{Kingma_2015} with a fixed learning rate of 0.001.
The mini-batch included 16 of the 4-second segments.

We then used a feature extractor of the SerialOE as $f_D(\cdot)$.
The network consisted of the EfficientNet B0~\cite{xie2020self}, 
global average pooling, and two nonlinear transformations, and it outputted a 128-dimensional feature.
Both $g_{\rm mac}(\cdot)$ and $g_{\rm id}(\cdot)$ were linear transformations, and $\lambda_{\rm id}$ was set to 5.
The audio input feature $\psi(\cdot)$ was the Mel-spectrogram with a window size of \SI{128}{\ms}, a hop size of \SI{16}{\ms}, and 224 Mel-spaced frequency bins in the range of \SIrange{50}{7,800}{\Hz}.
We trained the network for 150 epochs using the AdamW optimizer~\cite{adamw2019}, OneCycleLR scheduler~\cite{onecyclelr2019} with a learning rate of 0.001, and mixup~\cite{zhang2017mixup}.
The mini-batch included 128 of the 4-second segments and we used a batch sampler so that the ratio of the value of $t^{\rm mac}\in\{0,1\}$ was 1:1.
We also used the SerialOE as the comparison method and its $h$ was the GMM.

During inference, we divided the original 10-second signal into seven 4-second segments with \SI{75}{\percent} overlapping and calculated the final anomaly score by averaging the score of each segment.
As for the hyperparameters, the GMM of SerialOE included the number of clusters and the domain of its training data, and the proposed method included the number of neighbors $K$ and the domain of the test data used for smoothing.
These were selected based on the validation results and the resulting settings are shown in Table~\ref{tbl:hp}.
For reference, we also evaluated the Oracle version of the proposed method, which used the best hyperparameters on the evaluation set.

\begin{table}[t!]
\centering
\caption{Hyperparameters for each machine type. ``ALL'' of the ``Data'' columns means the use of both source and target domain.}
\label{tbl:hp}
\begin{tabular}{l|rl|rl|rl}
\toprule
 & \multicolumn{2}{l|}{SerialOE} & \multicolumn{2}{l|}{Proposed} & \multicolumn{2}{l}{Proposed (Oracle)} \\
 & \# clusters  & Data & $K$ & Data & $K$ & Data \\
\midrule
fan & 2 & Source & 512 & Source & 1,024 & Source\\
gearbox & 32 & Source & 1,024 & All & 2,048 & All\\
pump & 8 & Source & 128 & Source & 1,024 & Source\\
slider & 4 & All & 64 & Source & 128 & Source\\
ToyCar & 4 & All & 128 & Source & 2,048 & All\\
ToyTrain & 16 & Source & 1,024 & All & 256 & All\\
valve\ & 2 & All & 128 & Source & 1,024 & Source\\
\bottomrule
\end{tabular}
\vspace{-10pt}
\end{table}

\begin{table*}[t!]
\centering
\caption{Evaluation results.
Values of each machine represent the harmonic mean of the AUC~[\%] and pAUC ($p=0.1$)~[\%] over all sections.
Values of the All-hmean represent the harmonic mean of the AUC and pAUC over all machines and sections.
Values of the ToyCar-5 represent the AUC of ToyCar in section~5.
Hyperparameters of Proposed (Oracle) on the ToyCar-5 are 128 for $K$ and All for Data}
\label{tbl:eval}
\begin{tabular}{l|lllllll|l|l}
\toprule
Method & fan & gearbox & pump & slider & ToyCar & ToyTrain & valve & All-hmean & ToyCar-5\\
\midrule
SerialOE & \textbf{80.78} & 60.77 & \textbf{76.86} & \textbf{81.49} & 49.37 & 55.06 & 64.02 & 64.72 & 33.58\\
AE & 57.11 & 59.50 & 54.70 & 55.63 & 58.61 & 58.74 & 51.27 & 56.38 & 72.87\\
Proposed & 73.22 & 64.42 & 71.84 & 69.13 & 58.18 & 55.91 & 56.95 & 63.52 & 72.13\\
Proposed (Oracle)& 74.56 & \textbf{64.64} & 74.76 & 70.63 & \textbf{63.09} & \textbf{70.19} & \textbf{73.28} & \textbf{69.89} & \textbf{77.29}\\

\bottomrule
\end{tabular}
\vspace{-7pt}
\end{table*}

\subsection{Results}
The results in Table~\ref{tbl:eval} show that the proposed method \modified{improves} the performance of AE in machines other than ToyCar and ToyTrain, achieving an absolute improvement of \SI{7}{\percent} in All-hmean.
The proposed method \modified{makes} especially significant improvements larger than \SI{13}{\percent} in the fan, pump, and slider where SerialOE achieves high performance larger than \SI{75}{\percent}.
We can also see that SerialOE has an AUC less than \SI{50}{\percent} in ToyCar-5 and the discriminative approach has a risk that falls into significant performance degradation.
Even in such a situation, the proposed method \modified{works} robustly and even has the potential to improve the performance of original generative methods~(see Proposed (Oracle)) because an AUC significantly below \SI{50}{\percent} indicates that normal and anomalous sounds \modified{are} distinguished while the cluster of the anomalous sounds \modified{is} closer to $h$.
The proposed method can further improve its performance by setting proper hyperparameters (e.g., improvement of \SI{22}{\percent} in the valve).
Our future work includes its adaptive setting depending on the space.

\begin{figure}[t]
  \centering
  \centerline{\includegraphics[width=0.99\columnwidth]{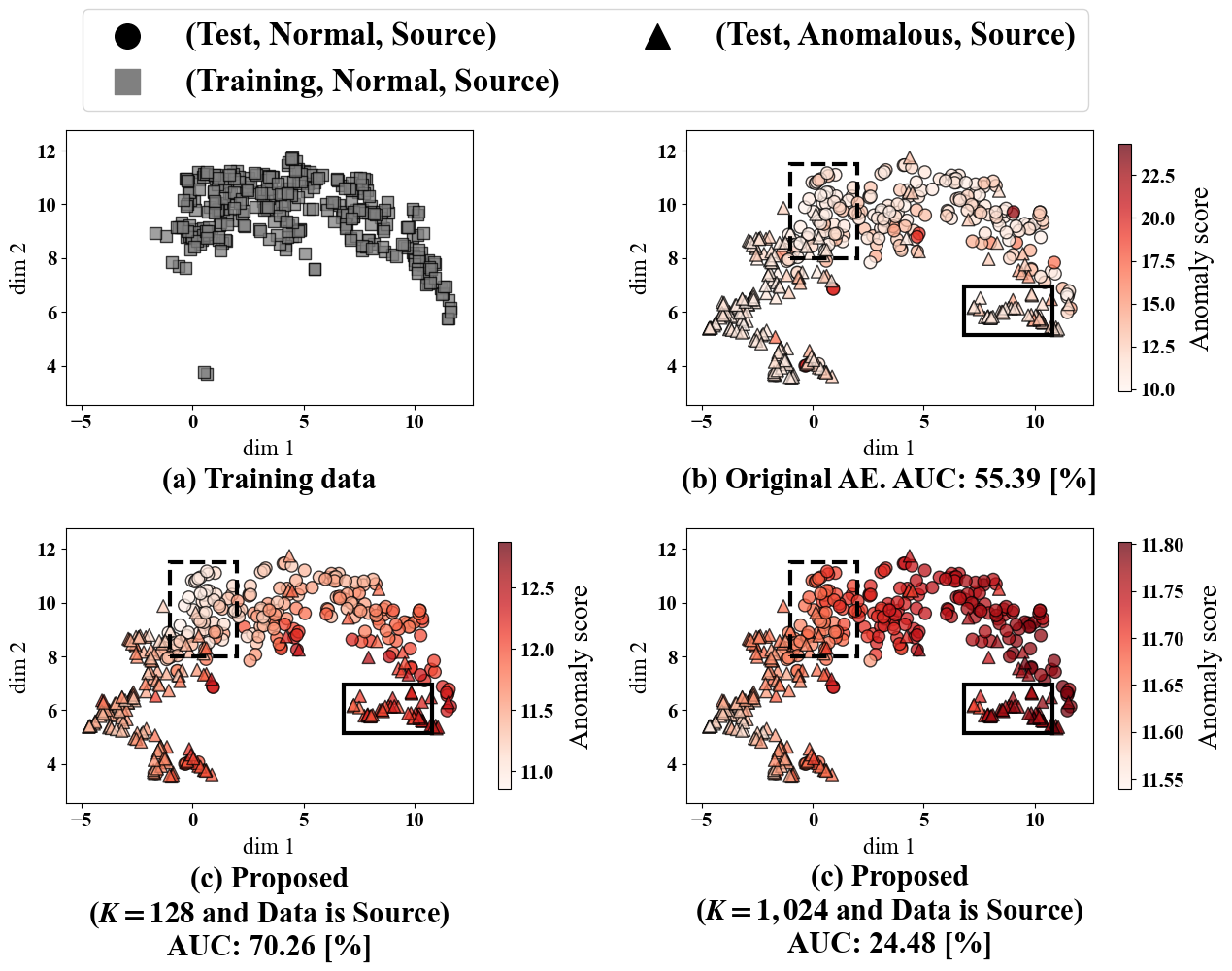}}
  \vspace{-5pt}
  \caption{Visualization of smoothing in the source domain of Section 4 of the slider.
  The colormap shows the anomaly score of $a_{\rm G}$ or $a_{\rm prop}$.
  For visibility, only \SI{30}{\percent} of the entire sample was randomly selected and displayed.
  All figures have the same axes.
  }
  \label{fig:slider}
  \vspace{-10pt}
\end{figure}

\begin{figure}[t]
  \centering
  \centerline{\includegraphics[width=0.99\columnwidth]{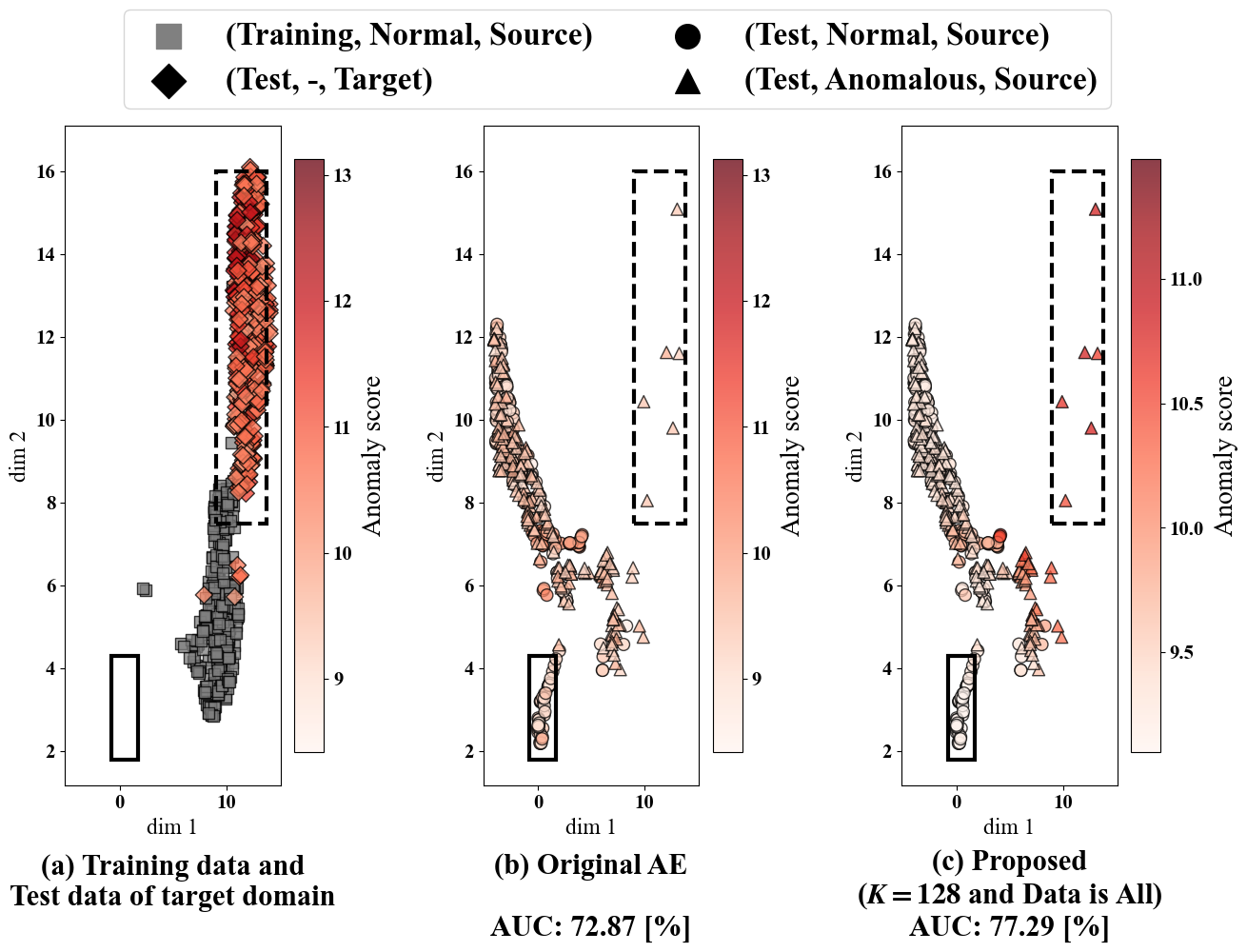}}
  \vspace{-5pt}
  \caption{Visualization of smoothing in Section 5 of ToyCar.
  The colormap shows the anomaly score of $a_{\rm G}$ or $a_{\rm prop}$.
  For visibility, only \SI{30}{\percent} of the entire sample was randomly selected and displayed.
  All figures have the same axes.}
  \label{fig:toycar}
  \vspace{-10pt}
\end{figure}

Figures~\ref{fig:slider} and \ref{fig:toycar} show the feature space using UMAP~\cite{mcinnes2018umap} with the original and smoothed anomaly scores.
In the solid frames in Fig.~\ref{fig:slider}, we can see that the proposed method \modified{modifies} the anomaly scores of the anomalous sounds to higher scores.
On the other hand, it is clear that a too-large $K$ causes performance degradation like in the dashed frame, where normal sound is also smoothed with anomalous sounds.
In the solid frames in Fig.~\ref{fig:toycar}, we can see that the normal sounds of the test data deviate from the cluster of the training data.
The proposed method robustly \modified{works} in such a situation and \modified{modifies} the anomaly scores of the normal sounds to lower scores.
In the dashed frames, anomalous sounds \modified{are} smoothed with the sounds of the target domain which have relatively high anomaly scores, resulting in performance improvements.

\begin{figure}[t]
  \centering
  \centerline{\includegraphics[width=0.99\columnwidth]{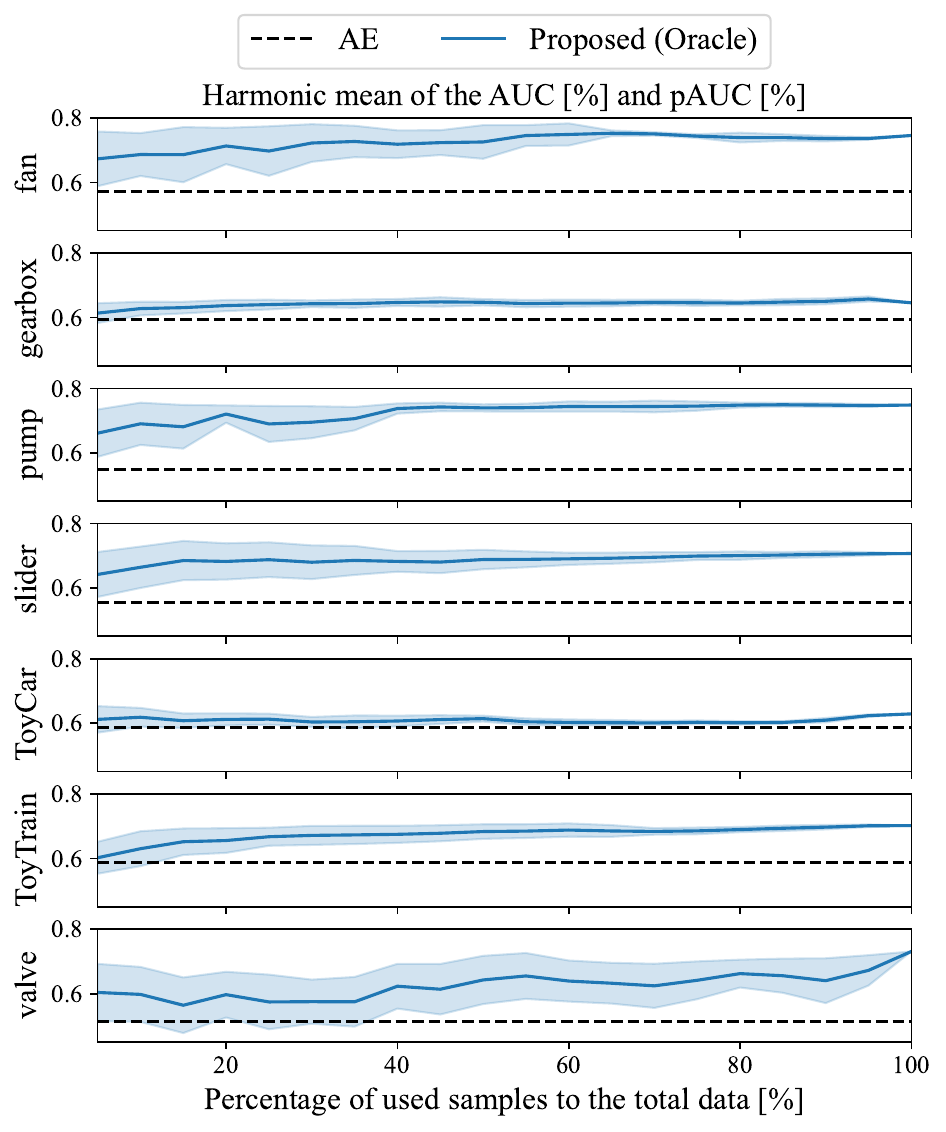}}
  \caption{Evaluation changing the number of samples used for smoothing. The evaluation metric is a harmonic mean of the AUC~[\%] and pAUC ($p=0.1$)~[\%] over all sections. The error band shows the standard deviation.}
  \label{fig:do}
  \vspace{-5pt}
\end{figure}

To investigate the behavior of the proposed method in a real situation where we can get new samples sequentially, we evaluate it by changing the number of samples used for smoothing.
Figure~\ref{fig:do} shows the evaluation result when the number of samples used for $\mathcal{Y}$ is changed from \SI{5}{\percent} to \SI{100}{\percent} of the total evaluation data in increments of \SI{5}{\percent}, while keeping the ratio of normal to anomalous and source to target domains in each section.
We randomly selected used samples and plotted the mean and standard deviation of 20 trials with different random seeds.
We always gave the best hyperparameter to the proposed method.
In Fig.~\ref{fig:do}, we can see an overall increasing trend in the AUC and a decreasing trend in the standard deviation corresponding to an increase in the number of samples.
The result also shows that even with the small number of samples, the proposed method can improve the original generative method by giving proper hyperparameters, although there is some variation.
This experiment shows that it is desirable to have regular updates of $\mathcal{Y}$; however, there is no additional cost for recording sounds, since ASD systems always do this for monitoring.
Rather, we believe that past data should be accumulated and utilized.

\section{Conclusion}
In this paper, we proposed the discriminative neighborhood smoothing of generative anomaly scores to realize high-performance ASD without the risk of degradation due to the discrepancy between the training and test data.
The proposed method improved the performance of the generative method by smoothing the anomaly scores of multiple samples with similar discriminative features.
This method does not suffer from the discrepancy because it does not use the inlier model to calculate the anomaly scores.
Our experimental results showed that
(1) the proposed method greatly improved the performance of AE,
(2) it robustly worked despite discrepancies,
(3) it had the potential to outperform SerialOE by removing the risk due to the discrepancy,
and (4) it is desirable to accumulate and utilize the past data for its stability.

\section*{Acknowledgment}
This paper was partly supported by a project, JPNP20006, commissioned by NEDO, JSPS KAKENHI Grant Number JP20H00102, and AIST KAKUSEI project (FY2023).


\section*{References}
\printbibliography

\end{document}

%% file: settings.tex
\usepackage[backend=biber,style=ieee,
maxbibnames=6,
doi=false,isbn=false,url=false,eprint=false
]{biblatex}
\defbibheading{bibliography}[\refname]{}

\DeclareSourcemap{
	\maps[datatype=bibtex, overwrite=true]{
		\map{
		    \step[fieldsource=booktitle,
			match=\regexp{.*EUSIPCO.*},
			replace={Proc. EUSIPCO}]
			\step[fieldsource=booktitle,
			match=\regexp{.*CVPR.*},
			replace={Proc. CVPR}]
			\step[fieldsource=booktitle,
			match=\regexp{.*Interspeech.*},
			replace={Proc. Interspeech}]
			\step[fieldsource=booktitle,
			match=\regexp{.*ICASSP.*},
			replace={Proc. ICASSP}]
			\step[fieldsource=booktitle,
			match=\regexp{.*ICLR.*},
			replace={Proc. ICLR}]
			\step[fieldsource=booktitle,
			match=\regexp{.*ICML.*},
			replace={Proc. ICML}]
			\step[fieldsource=booktitle,
			match=\regexp{.*ASRU.*},
			replace={Proc. ASRU}]
			\step[fieldsource=booktitle,
			match=\regexp{.*SLT.*},
			replace={Proc. SLT}]
			\step[fieldsource=booktitle,
			match=\regexp{.*SSW.*},
			replace={Proc. SSW}]
			\step[fieldsource=booktitle,
			match=\regexp{.*WASPAA.*},
			replace={Proc. WASPAA}]
            \step[fieldsource=booktitle,
            match=\regexp{.*Detection.*and.*Classification.*of.*Acoustic.*Scenes.*and.*Events.*Workshop.*},
			replace={Proc. DCASE}]
            \step[fieldsource=booktitle,
			match=\regexp{.*MLSP.*},
			replace={Proc. MLSP}]
            \step[fieldsource=booktitle,
			match=\regexp{.*ECCV.*},
			replace={Proc. ECCV}]
            \step[fieldsource=journal,
			match=\regexp{.*NeurIPS.*},
			replace={Advances in NuerIPS}]
            \step[fieldsource=journal,
			match=\regexp{.*TASLP.*},
			replace={IEEE/ACM TASLP}]
            \step[fieldsource=journal,
			match=\regexp{.*J-STSP.*},
			replace={IEEE J-STSP}]                
			\step[fieldsource=series,
			match=\regexp{.+},
			replace={{}}]
			\step[fieldsource=editor,
			match=\regexp{.+},
			replace={{}}]
			\step[fieldsource=publisher,
			match=\regexp{.+},
			replace={{}}]
			\step[fieldsource=month,
			match=\regexp{.+},
			replace={{}}]
			\step[fieldsource=location,
			match=\regexp{.+},
			replace={{}}]
			\step[fieldsource=address,
			match=\regexp{.+},
			replace={{}}]
			\step[fieldsource=organization,
			match=\regexp{.+},
			replace={{}}]
		}
	}
}